\documentclass[pre,twocolumn]{revtex4-1}
\usepackage{subcaption}
\usepackage{graphicx}
\usepackage{comment}
\usepackage{amsmath,amssymb,amsthm} 
\usepackage{scalerel,stackengine}
\usepackage{bm}
\usepackage{verbatim}
\usepackage{float}

\begin{document}
	\title{  The energy, temperature, and heat capacity in  discrete classical dynamics}

\author{ S\o ren  Toxvaerd }
\affiliation{ Department of Science and Environment, Roskilde University, Post box 260, DK-4000 Roskilde, Denmark}
\email{st@ruc.dk}

\begin{abstract}
 Simulations  of objects with classical dynamics are in fact a particular version of discrete  dynamics
since almost all the classical dynamics simulations in natural science are performed with the use of the simple ''Leapfrog" or ''Verlet" 
 algorithm. It was, however, Newton who in $Principia$, $ Proposition$ $I$ in 1687 first
formulated the discrete algorithm, which much later in 1967 was rederived by L. Verlet. Verlet also formulated a first-order approximation
	for the velocity $\textbf{v}(t)$
	at time $t$, which has been used in  simulations since then. The approximated expressions 
	for  $\textbf{v}(t)$ and the kinetic energy lead to
	severe errors in the thermodynamics at high densities, temperatures, strong repulsive forces, or for large discrete time increments
	used in the discrete ''Molecular Dynamics" (MD) simulations. Here we derive the exact expressions for the discrete dynamics,
	and show by simulations of
	a Lennard-Jones system, that these expressions now 
	result in equality between temperatures determined from the kinetic energies, and the corresponding configurational
	temperatures determined from the expression of Landau and Lifshitz, derived from the forces.
\\

\doi 10.1103/PhysRevE.00.005300

\end{abstract}
\maketitle

\section{Introduction}
 Simulations  of objects with classical dynamics are in fact a particular version of discrete  dynamics
since almost all the classical dynamics simulations in natural science are performed with the use of a 
simple algorithm, first formulated by Newton \cite{Newton1687,Toxvaerd2020}.
   A new  position in Newton's  discrete dynamics, $\textbf{r}_i(t+\delta t)$, at time $t+\delta t$ of an object  
$i$ with the mass $m_i$   is determined by
the force $\textbf{f}_i(t)$ acting on the object   at the discrete position $\textbf{r}_i(t)$  at time $t$ together with 
 the position $\textbf{r}_i(t-\delta t)$ at $t - \delta t$  as
\begin{equation}
	 m_i\frac{\textbf{r}_i(t+\delta t)-\textbf{r}_i(t)}{\delta t}
			=m_i\frac{\textbf{r}_i(t)-\textbf{r}_i(t-\delta t)}{\delta t} +\delta t \textbf{f}_i(t),	
 \end{equation}
where the velocities  $ \textbf{v}_i(t+\delta t/2) =  (\textbf{r}_i(t+\delta t)-\textbf{r}_i(t))/\delta t$ and
 $  \textbf{v}_i(t-\delta t/2)=  (\textbf{r}_i(t)-\textbf{r}_i(t-\delta t))/\delta t$ and corresponding momenta are constant in
the time intervals in between the discrete positions.
Newton  begins \textit{Principia} by postulating Eq. (1) in \textit{Proposition I},
and he obtained his second   law   as the  limit $ lim_{\delta t \rightarrow 0}$ of the equation.
For Newton's  derivation of his second law and his discrete algorithm in \textit{Proposition I} see Ref. 3, Chap. 2 \cite{Toxvaerd2023}.

  The algorithm, Eq. (1), is usually  presented  as the Leapfrog algorithm 
\begin{eqnarray}
\textbf{v}_i(t+\delta t/2)=  \textbf{v}_i(t-\delta t/2)+ \delta t/m_i  \textbf{f}_i(t)  \\
\textbf{r}_i(t+\delta t)= \textbf{r}_i(t)+ \delta t \textbf{v}_i(t+\delta t/2),	  
\end{eqnarray}	  
where the positions are obtained from the discrete values of the velocities.
The rearrangement of Eq. (1) gives  the Verlet algorithm \cite{Verlet1967}

\begin{equation}
	\textbf{r}_i(t+\delta t)=2\textbf{r}_i(t)-\textbf{r}_i(t-\delta t) +\delta t^2 \textbf{f}_i(t)/m_i .	  
\end{equation}	

The algorithm is used in almost all discrete ''Molecular Dynamics'' (MD) simulations and with Verlet's expression for the velocity at the 
discrete time $t$
\begin{equation}
	\textbf{v}_0(t)=\frac{\textbf{r}(t+\delta t) -\textbf{r}(t-\delta t)}{2\delta t}.
\end{equation}

 The approximation (5) for the velocity $\textbf{v}(t)$ is the first term in a  symmetrical Taylor expansion from the position \textbf{r}(t) on the classical analytic trajectory,
but the discrete dynamics trajectory is not analytic. Furthermore, there are times for an analytic trajectory where a
Taylor expansion is very slowly converging, e.g. at particle collisions or for fast vibrations of atoms in molecules
with covalent bonds.
The velocity is, however, as Newton noticed in $Principia$ constant in times except at the discrete times where it is exposed to
a force impulsive (see next section), and it is given by Eq. (2). Verlet's inaccurate expression for the velocity results in a corresponding
inaccurate measure of the energy, temperature, and  heat capacity.

The purpose of the article is to show that Newton's discrete dynamics has the same properties as his  analytic Classical Mechanics, and
to correct the expressions
for the energy,  temperature,  and heat capacity in MD simulations \cite{Toxvaerd2023}.

In  Sec. II we first present Newton and Verlet's formulation of the discrete classical dynamics and 
derive the energy invariance. The connections between the ''Verlet'' expressions for the energy, temperature, and
heat capacity and the corresponding exact Newtonian discrete  dynamic expressions are derived in Section III.
In Section IV we determine the differences between the traditional (Verlet)  values of the temperature
  and heat capacities, and the corresponding Newtonian values
for a system of $N=2000$ particles with Lennard-Jones (LJ) interactions. Section V summarizes the results.

\section{Discrete Newtonian dynamics}
  The classical discrete dynamics between $N$ spherically symmetrical objects
     with masses $ m^N=m_1, m_2,..m_i,..,m_N$ and positions \textbf{r}$^N(t)=$\textbf{r}$_1(t)$, \textbf{r}$_2(t)
     ,..,$\textbf{r}$_i(t),..$\textbf{r}$_N(t)$  is obtained 
 by Eq. (1). Let the force $ \textbf{f}_i(t)$ on object  $i$ be a sum of pairwise  forces  $ \textbf{f}_{ij}(t)$ between pairs of   objects $i$ and $j$
 \begin{equation}
	 \textbf{f}_i(t)=  \sum_{j \neq i}^{N} \textbf{f}_{ij}(t).
 \end{equation}	

\subsection{Newton's formulation of discrete classical dynamics}

Newton  begins \textit{Principia} by postulating Eq. (1) in \textit{Proposition I}. The English translation
of \textit{Proposition I} is

\textit{Of the Invention of Centripetal Forces.}\\
  PROPOSITION I. Theorem I.\\
 \textit{The areas, which revolving bodies describe by radii drawn to an immovable centre of force do lie in the same immovable planes, and
 are proportional to the times in which they are described}.\\
\textit{ For suppose the time to be divided into equal parts, and in the first part of time let the body by its innate force describe the right line
AB. In the second part of that time, the same would (by Law I.), if not hindered, proceed directly to c, along the line Bc equal to AB; so that the radii
AS, BS, cS, drawn to the centre, the equal areas ASB, BSc, would be described. But when the body is arrived at B,
\textbf{suppose that a centripetal force acts at once
with a great impulse}, and, turning aside the body from the right line Bc, compels it afterwards
to continue its motion  along the right line BC. Draw cC parallel
to BS meeting BC in C; and at the end of  the second part of the time, the body (by Cor. I of Laws) will be found in C, in the same plane with the
triangle ASB. Join SC, and,  because SB and Cs are parallel, the triangle SBC will be equal to the triangle SBc, and therefore also to the
triangle SAB. By the like argument, if the centripetal force acts successively in C, D, E, \& c., and makes the body,
in each single particle of time, to describe the right lines CD, DE, EF, \& c., they will all lie in the same plane;
and the triangle SCD will be equal to the triangle SBC, and SDE to SCD, and SEF to SDE. And therefore, in equal times, equal areas are described in on immovable plane:
and, by composition, any sums SADS, SAFS, of those areas, are one to the other as the times in which they are described. Now let the number of
those triangles be augmented; and their breadth diminished in infinitum; and (by Cor. 4, Lem III) their ultimate perimeter ADF will be a curve line:
and therefore the centripetal force, by which the body is perpetually drawn back from the tangent of this curve, will act continually; and any described
areas SADS, SAFS, which are always proportional to the times of description, will, in this case also, be proportional to those times.} Q. E. D.

$\textit{Proposition I}$ in $\textit{Principia}$ is illustrated with  a figure (Fig. 1).
The assumption ..\textit{\textbf{suppose that a centripetal force acts at once with a great impulse}}..
in \textit{Proposition I}  is highlighted here. It is the central part in $Principia$. Newton's
second law is derived from his discrete dynamics, where the time, forces, and the change in momenta are quantized and
the velocities are constant in between the discrete times. Newton's discrete dynamics in  \textit{Proposition I}
is in fact Classical Quantum Dynamics \cite{Toxvaerd2023}.

Nobel Laureate T. D. Lee was perhaps the first to suggest that difference equations
are to be preferred in the
foundation of dynamics, and with a classical difference equation as the  classical limit path of Feynman's quantum paths \cite{Lee1987}. This classical limit path is Newton's
discrete dynamics in \textit{Proposition I}.
Our classical world is of course analytic. The difference between analytic and discrete dynamics is 
 proportional to the square of $\delta t=$ Planck time ($\approx 10^{-43}$ sec), and negligible.
 But one can say that  Newtonian discrete dynamics simulations with time increments
 $\delta t \approx 10^{-14}$ sec, used in many MD simulations   is in the same ''universality class'' as the analytic dynamics,
 and exact in the same manner as classical analytic dynamics.
If so it places Newtonian discrete dynamics as a remarkably useful scientific tool.
\begin{figure}
\begin{center}	  
\includegraphics[width=8.6cm,angle=0]{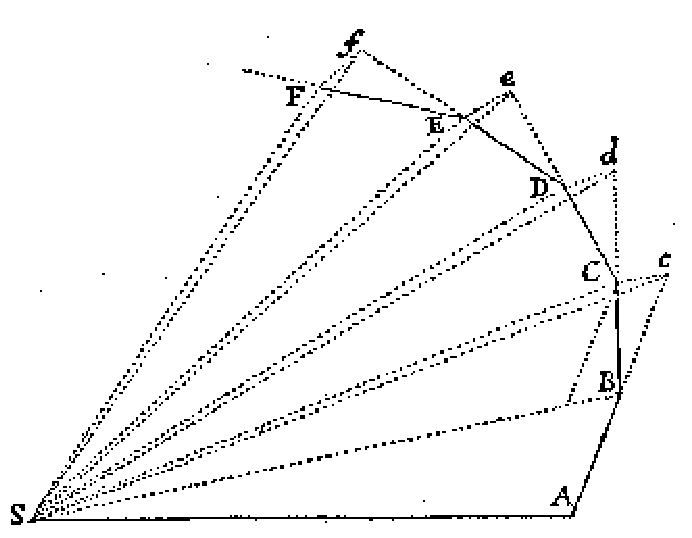}
\caption{Newton's figure at $Proposition$ $I$ in Principia, with the formulations of the discrete dynamics. 
	  The discrete positions are A: $\bf{r}$$(t-\delta t)$;  B: $\textbf{r}(t)$;  C: $\textbf{r}(t+\delta t)$, etc.. The deviation cC
	  from the straight line ABc (Newton's first law) is caused by a  force from  the position S at time $t$.}
\end{center}		  
\end{figure}

\subsection{The Verlet algorithm}

Loup Verlet (1931-2019) published in 1967 the article \textit{ Computer ''Experiments" on Classical Fluids. I. Themodynamical
Properties of Lennard-Jones Molecules} \cite{Verlet1967}, where Newton's discrete algorithm with the formulation by Eq (4)
was introduced. The algorithm was derived by
a forward and backward Taylor expansion \cite{Toxvaerdappendix}
\begin{eqnarray}
\textbf{r}(t+\delta t)= \textbf{r}(t)+ \delta t \frac{\partial \textbf{r}(t)}{\partial t}
+\frac{1}{2} \delta t^2 \frac{\partial^2 \textbf{r}(t)}{\partial t^2}+
	\frac{1}{6}\delta t^3\frac{\partial^3 \textbf{r}(t)}{\partial t^3}+.. \nonumber 	\\
\textbf{r}(t-\delta t)= \textbf{r}(t)- \delta t \frac{\partial \textbf{r}(t)}{\partial t}
+\frac{1}{2} \delta t^2 \frac{\partial^2 \textbf{r}(t)}{\partial t^2}-
	\frac{1}{6}\delta t^3\frac{\partial^3 \textbf{r}(t)}{\partial t^3}+..,
\end{eqnarray}
and the algorithm was obtained from  the sum  $ \textbf{r}(t+\delta t)+\textbf{r}(t-\delta t)$ and
$ \delta t^2 \partial^2 \textbf{r}(t)/\partial t^2=\frac{\delta t^2}{m} \textbf{f}(t)$. All the odd
terms in the sum cancel, and the Verlet algorithm was formulated as a 
four-order time symmetric predictor of the positions at the analytic trajectories.
The velocity is correspondingly derived from the difference $\textbf{r}(t+\delta t)-\textbf{r}(t-\delta t)$, 
all even terms cancel, an it is

\begin{equation}
	\textbf{v}(t)=\frac{\textbf{r}(t+\delta t)-\textbf{r}(t-\delta t)}{ 2 \delta t}
	+\frac{1}{6}\frac{\delta t^2}{m} \textbf{f}'(t)+.. 
\end{equation}

The scientific community and Verlet were
first much later aware, that it
actually was Newton who first published the geometric formulation of the algorithm in $Proposition$ $I$ \cite{Nauenberg2018a}.

\subsection{The energy invariance in discrete Newtonian dynamics}

  Newton's algorithm is a  symmetric time centered difference whereby the dynamics is time reversible and symplectic.
  The conservation of momentum and angular momentum for a conservative system follows directly from Newton's third law  for the
  conservative system with the forces $\textbf{f}_{ij}(t)=-\textbf{f}_{ji}(t)$
  between objects  $i$ and $j$, but the energy invariance is not so obvious.

 The  energy in a system with analytic dynamics is the sum of potential energy
 $ U(\textbf{r}^N(t))$ and kinetic energy $K(t)$, 
 and it is  time invariant for a conservative system.
 The kinetic energy  in the discrete dynamics at time $t$ is, however, not well-defined since the velocities change   at  $t$. 
Traditionally one uses Verlet's  first-order expression for the velocity at time $t$, Eq. (5),
obtained by the symmetric Taylor expansion, and

 \begin{eqnarray}
	 K_0(t)=  \sum_i^N \frac{1}{2}m_i \textbf{v}_{0,i}(t)^2 \\
	E_0(t)= U(\textbf{r}^N(t))	+K_0(t)
 \end{eqnarray}	 
  The energy $E_0$ obtained by using  the approximation  (5) and with $K(t)=K_0(t)$ for the kinetic energy 
  fluctuates with time, although it is constant when averaged over long time intervals.

  The velocities are, however, constant in between the discrete times in Newton's discrete dynamics, and
the energy invariance can be obtained by considering the energy in
the time interval $[t- \delta t/2, t+ \delta t/2]$, and dividing the interval into two
sub-intervals $[t-\delta t/2,t]$ and $[t, t+\delta t/2]$.  
The energy invariance, $E_{\textrm{D}}$ in Newton's discrete dynamics (D)
 can then be  obtained by considering the change in kinetic energy, $\delta K_ {\textrm{D}},$
 the change in potential energy,  $\delta U_ {\textrm{D}},$ and
the work,  $W_{\textrm{D}}$ done by the forces 
in the time interval $[t-\delta t/2, t+\delta t/2].$

The loss in  potential energy, $-\delta U_{\textrm{D}}$ is defined as
the work done by the forces for a change of  positions \cite{Goldstein}. 
The work, $W_{\textrm{D}}$ done in the time interval by the discrete dynamics
from the position  $(\textbf{r}_i(t)+ (\textbf{r}_i(t-\delta t))/2$ at $t-\delta t/2$
to   the position  $(\textbf{r}_i(t+\delta t)+ \textbf{r}_i(t))/2$ at $t+\delta t/2$,
and with the change $\delta \textbf{r}_i$ of the position
$ \delta \textbf{r}_i= \textbf(\textbf{r}_i(t+\delta t) -\textbf{r}_i(t-\delta t))/2$ is  \cite{Toxvaerd2023}
\begin{equation}
	-\delta U_{\textrm{D}}=W_{\textrm{D}}= \sum_i^N  \textbf{f}_i(t)(\textbf{r}_i(t+\delta t) -\textbf{r}_i(t-\delta t))/2.
\end{equation}	
By rewriting Eq. (4) to
\begin{equation}
	\textbf{r}_i(t+ \delta t) -\textbf{r}_i(t-\delta t)= 2(\textbf{r}_i(t) -\textbf{r}_i(t-\delta t))+\frac{\delta t^2}{m_i} \textbf{f}_i(t),
\end{equation}
and inserting in Eq. (11) one obtains an expression for the total work in the time interval
\begin{equation}
	-\delta U_{\textrm{D}}=  W_{\textrm{D}}	=  \sum_i^N  [(\textbf{r}_i(t) -\textbf{r}_i(t-\delta t)) \textbf{f}_i(t)+ \frac{\delta t^2}{2m_i}\textbf{f}_i(t)^2].
\end{equation}

The mean kinetic energy $K_{\textrm{D}}$ of the discrete dynamics  in the time interval $[t-\delta t/2, t+\delta t/2]$ is

\begin{eqnarray}	
 K_{\textrm{D}}=                           \nonumber \\		
\frac{1}{2} \sum_i^N \frac{1}{2}m_i[\frac{\textbf(\textbf{r}_i(t+\delta t/2)-\textbf{r}_i(t))^2}{\delta (t/2)^2}+
\frac{\textbf(\textbf{r}_i(t)-\textbf{r}_i(t-\delta t/2))^2}{\delta (t/2)^2}] \nonumber \\
=\frac{1}{2}\sum_i^N\frac{1}{2}m_i [\frac{\textbf(\textbf{r}_i(t+\delta t)-\textbf{r}_i(t))^2}{\delta t^2}+
\frac{\textbf(\textbf{r}_i(t)-\textbf{r}_i(t-\delta t))^2}{\delta t^2}],
\end{eqnarray}
 with the change 
\begin{eqnarray}	
\delta K_{\textrm{D}}=                           \nonumber \\		
\sum_i^N\frac{1}{2}m_i [\frac{\textbf(\textbf{r}_i(t+\delta t)-\textbf{r}_i(t))^2}{\delta t^2}-
\frac{\textbf(\textbf{r}_i(t)-\textbf{r}_i(t-\delta t))^2}{\delta t^2}].
\end{eqnarray}

By rewriting  Eq. (4) to
\begin{equation}
	\textbf{r}_i(t+ \delta t) -\textbf{r}_i(t)= \textbf{r}_i(t) -\textbf{r}_i(t-\delta t)+\frac{\delta t^2}{m_i} \textbf{f}_i(t)
\end{equation}
   and inserting the squared expression for  $\textbf{r}_i(t+\delta t) -\textbf{r}_i(t)$ in  Eq. (15), the change in kinetic energy
   is
\begin{equation}
	\delta K_{\textrm{D}}= \sum_i^N [ (\textbf{r}_i(t)-\textbf{r}_i(t - \delta t))\textbf{f}_i(t) +\frac{\delta t^2}{2m_i} \textbf{f}_i(t)^2].
\end{equation}

	 \begin{figure}
	  \begin{center}	  
 	 \includegraphics[width=5.6cm,angle=-90]{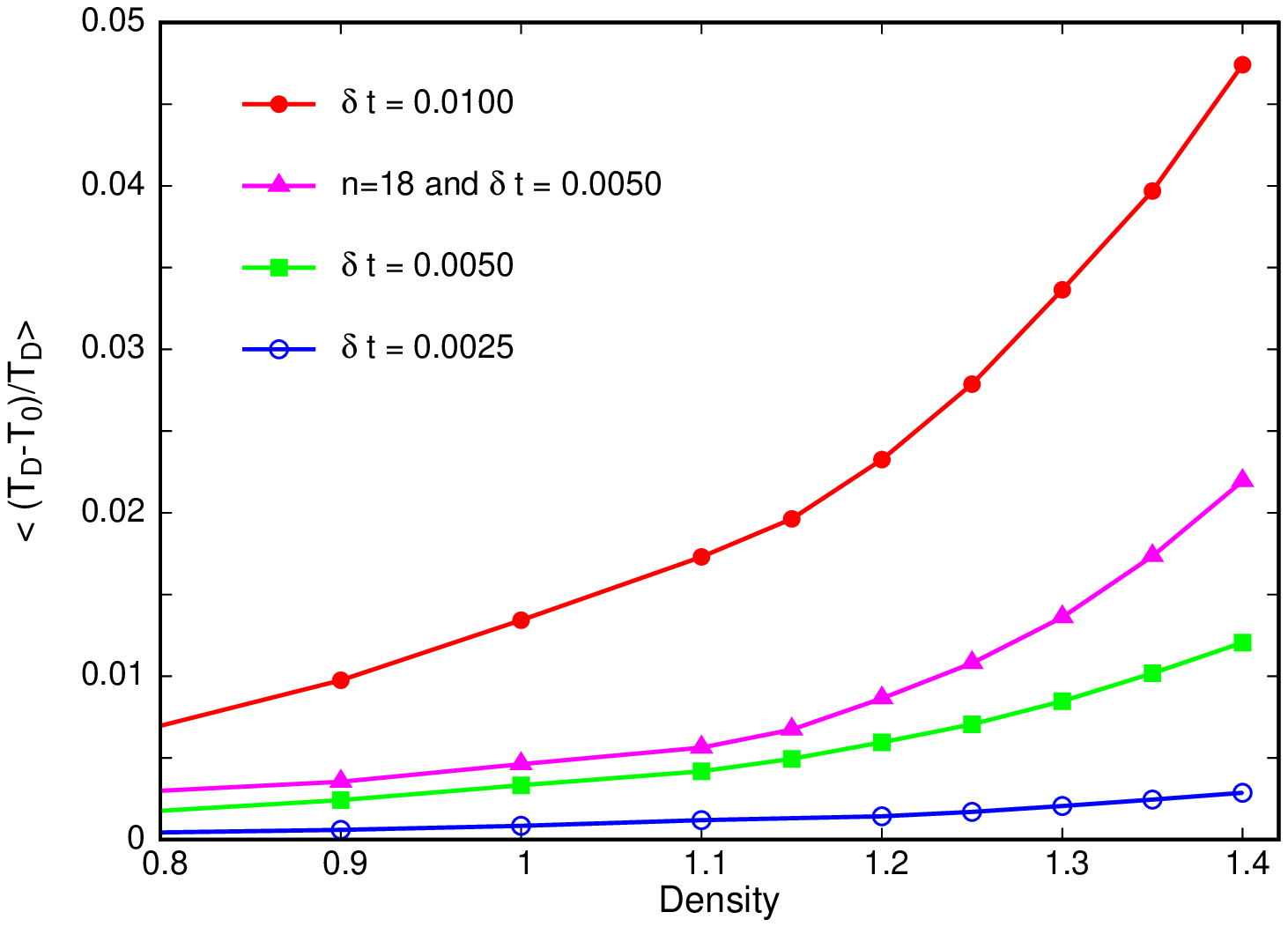}
		  \caption{  The relative mean difference $<(T_D-T_0)/T_D>$ between $T_D$ and $T_0$  as a function of the density
		  for a LJ fluid system at $T_D=1$. }
	  \end{center}		  
  \end{figure}

The energy invariance from the a discrete change of time from $t-\delta t/2$ to  $t+\delta t/2$ 
in Newton's discrete dynamics is expressed by Eqn. (13),  and  Eq. (17) as \cite{Toxvaerd2023}
\begin{equation}
	\delta E_{\textrm{D}}=	\delta U_{\textrm{D}}+\delta K_{\textrm{D}}=0.
\end{equation}

\section{The kinetic energy, temperature, and heat capacity in a Newtonian discrete conservative system}
 The constant velocities in the time intervals in between a force impulse  at time $t$ are related.
 It is easy to derive the relation  \cite{Toxvaerd2013}
\begin{equation}
	\textbf{v}_{0,i}(t)^2=\frac{1}{2}\textbf{v}_i(t+\delta t/2)^2+\frac{1}{2}\textbf{v}_i(t-\delta t/2)^2-
	\frac{1}{4}(\frac{\delta t}{ m_i }\textbf{f}_i(t))^2
\end{equation}
 from Eq. (4) between the square of the first term in Verlet's Taylor expansion  $\textbf{v}_{0,i}(t)^2$ , Eq. (5),
 for the  velocity at time $t$  and Newton's exact  expression for the  square of the
 constant velocities in the time intervals $[t-\delta t, t]$ and  $[t, t+\delta t]$.
The corresponding expression for the kinetic energy, $K_0(t)$, and the traditional value for the temperature used in MD simulations
\begin{equation}
	T_0=\frac{<2 K_0(t)>}{N_f}
\end{equation}	
 used in MD simulations
	for a system with $N_f$ degrees of freedom is less than the  mean kinetic energy, $K_D$, and the temperature $T_{\textrm{D}}$. In the
discrete time interval   $[t-\delta t/2, t+\delta t/2]$  the relation is
\begin{equation}
	K_0(t)=K_D(t)-\sum_i^N \frac{1}{8}\frac{\delta t^2}{ m_i} \textbf{f}_i(t)^2,
\end{equation}
and the systematic difference $T_D-T_0$ in the time interval by using Verlet's first-order approximation is
\begin{equation}
	T_D(t)-T_0(t)=\sum_i^N \frac{1}{4}\frac{\delta t^2}{ m_i} \textbf{f}_i(t)^2/N_f.
\end{equation}

 Newton's discrete dynamics  depends purely on the positions and the forces. The momenta are not dynamic variables, but  only 
 ''book-keeping" expressions in the dynamics \cite{Toxvaerd2023}. Therefore, the configurational temperature
\begin{equation}
	T_{\textrm{Conf}}(t)= <\nabla^2 U(\textbf{r}^N)>=
	\frac{\sum_i^N \textbf{f}_i(t)^2}{-\sum_i^N \nabla \cdot \textbf{f}_i(t)}.
\end{equation}
 could be a more relevant expression for the temperature
since it depends purely on the forces.  Eq. (23) is derived from the  average  of the Laplacian of the potential energy $\nabla^2  U(\textbf{r}^N)$, and
 is obtained from canonical averaging in the configurational phase space \cite{Landau,Rugh1997}.
The value of $T_{\textrm{Conf}}(t)$ 
 fluctuates with time, but its mean value obtained from long  simulations agrees with the corresponding temperatures,
$T_D$ obtained from the kinetic energy.

 The heat capacity  $C_V= 3/2R+C_V^i$ consists of two terms, the term from the kinetic energy and the term $C_V^i$
 from the interactions. The heat capacity can be determined either from numerical differentiation of
 the energy $U(T,\rho)$ or from the
  mean square fluctuation $< K(t)^2> -\bar{K}^2$ of the kinetic energy \cite{Lebowitz1967}.
The term  $K_D(t)-K_0(t)=\sum_i^N \frac{\delta t^2}{8m}\textbf{f}_i^2 $  in the  kinetic energy $K_0(t)$ is important
at high densities with hard particle collisions, and it affects
  $C_V^i$. The ratio
 \begin{equation}
\Delta C_V^i=	 \frac{	< (K_D(t) - \bar{K}_D)^2>-<(K_0(t) -\bar{K}_0)^2>}{ <(K_D(t)-\bar{K}_D)^2>}
 \end{equation}
 is a measure of the relative error in  $C_V^i$  by using Eq. (9) for the kinetic energy.
 \begin{figure}
	  \begin{center}	  
 	 \includegraphics[width=5.6cm,angle=-90]{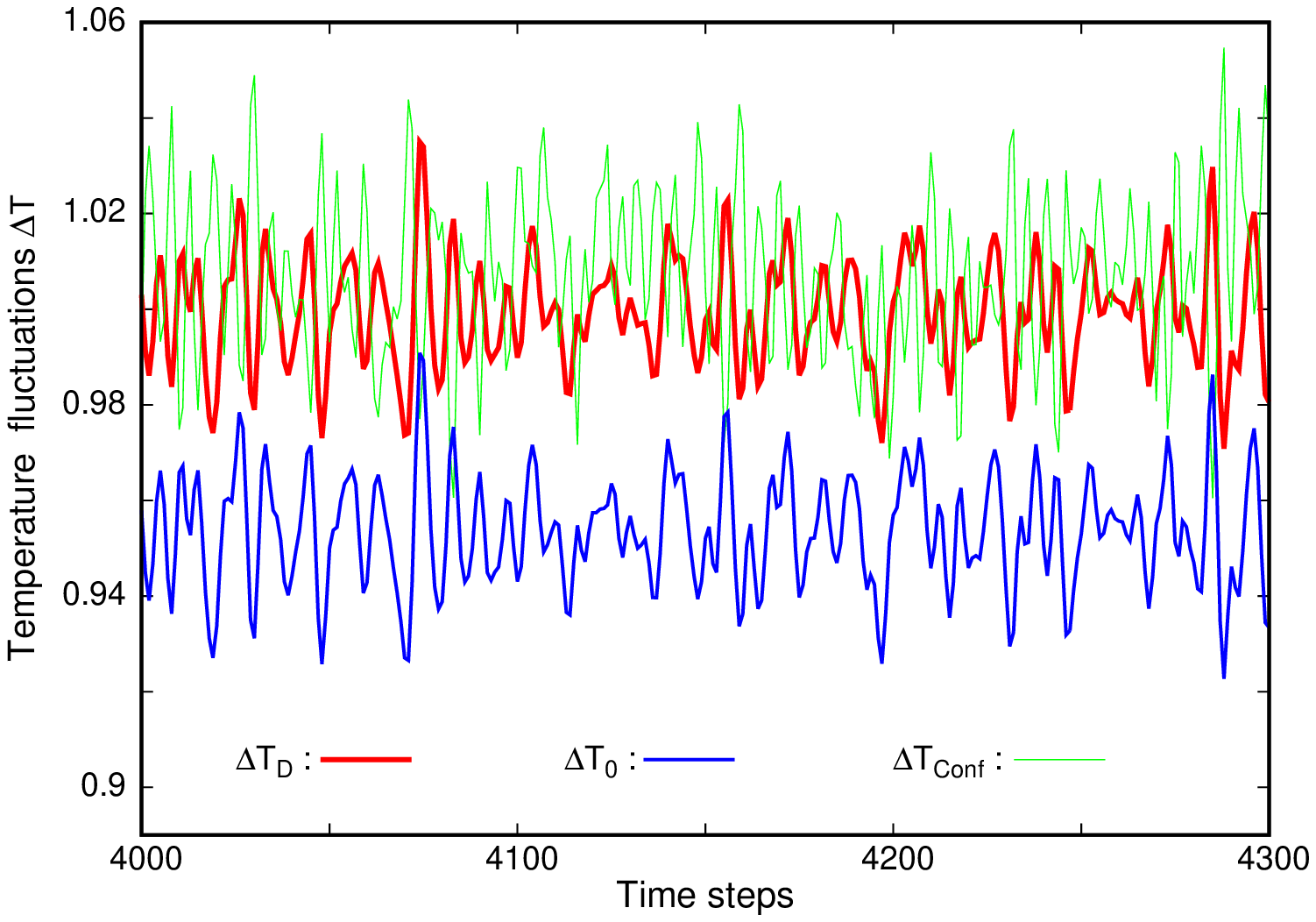}
		  \caption{ Temperature fluctuations in a $NVE$ system of $N=2000$ LJ particles at the density $\rho=1.40$, temperature
		  $T_D=1$, and with $\delta t=$0.010. With red is $T_D(t)$; with blue is the traditional expression $T_0(t)$ for the
		  temperature, and with thin green is the configurational temperature $T_{Conf}(t)$.
  }
	  \end{center}		  
  \end{figure}
\section{Simulations.}
Verlet used  Lennard-Jones (LJ) forces to simulate argon and later his  force field (FF) has been a standard  used in many 
simulations, e.g., simulations of  organic molecules to describe the  atom-atom FF between the atoms in  the molecules together with
 stronger  FF, e.g. for covalent bonds.
The errors $T_D-T_0$,  $E_D-E_0$, and in the heat capacities  by using Verlet's first order approximation for the velocity are biggest for 
strong force fields, and especially at collisions where the  FF is strongly repulsive. So the errors for systems  with strong
intramolecular FF and by  using Eq. (20) are
 bigger than for an LJ system.  

 LJ systems  with $N=2000$ LJ particles were simulated  at $T_D=1.00$ for different densities,
temperatures, strength of the repulsive forces, and time increments.
The difference between $T_D$ and $T_0$  at $T_D=1.00$ as a function of density for  $10^6$ 
time steps are shown in Fig. 2 for different time increments $\delta t$ and
repulsive forces given by the exponent $n$ in the $n-6$ LJ potential \cite{Heyes1998} . The differences increase
with density, temperature, the strength of the repulsive forces, or the discrete time step $\delta t$, but they are
relatively small  at state points with low densities. But the differences are significant for state points
with high densities,  systems with stronger  forces, and for large time increments $\delta t$.

 Most MD simulations are with a thermostat ($NVT$ and $NPT$ ensemble MD) and with the thermostat temperature $T(thermostat)=T_0$. We
 have performed $NVT$ simulations with a Nos\'{e}-Hover thermostat \cite{Hoover,Toxvaerd2024} and with
 $T(thermostat)=T_0$ as well as  $T(thermostat)=T_D$.
 The results from the $NVT$  simulations agree with the $NVE$ data. So the $NVE$ results in Fig. 2
  are also valid for the other ensemble simulations.

The fluctuations of the  temperatures $T_D(t), T_0(t)$, and $T_{Conf}(t)$ in the time interval $t \in [4000 \delta t, 4300 \delta t]$
for the MD  are shown in Fig. 3.
The temperatures are obtained for $N=2000$ LJ particles at $(T_D,\rho)=(1.00, 1.40)$, and with $\delta t= 0.010$.
The red curve is  for three hundred time steps  for $T_D(t)$, and the green curve is  $T_{Conf}(t)$. 
The mean temperatures  $T_D$ and $T_{Conf}$ from $10^6$ time steps  are approximately equal,
 $T_D=   0.999 \pm   0.011$ and $T_{Conf}=  1.008 \pm   0.020$, and higher than
the mean 
 $ T_0= 0.954 \pm 0.011$  for $T_0(t)$ shown in blue.
The temperatures in Fig. 3 show, that the temperature $T_D$ in Newtonian discrete dynamics is consistent with the otherwise obtained configurational temperature
$T_{Conf}$, but is not consistent with $T_0$.

	 \begin{figure}
	  \begin{center}	  
 	 \includegraphics[width=5.6cm,angle=-90]{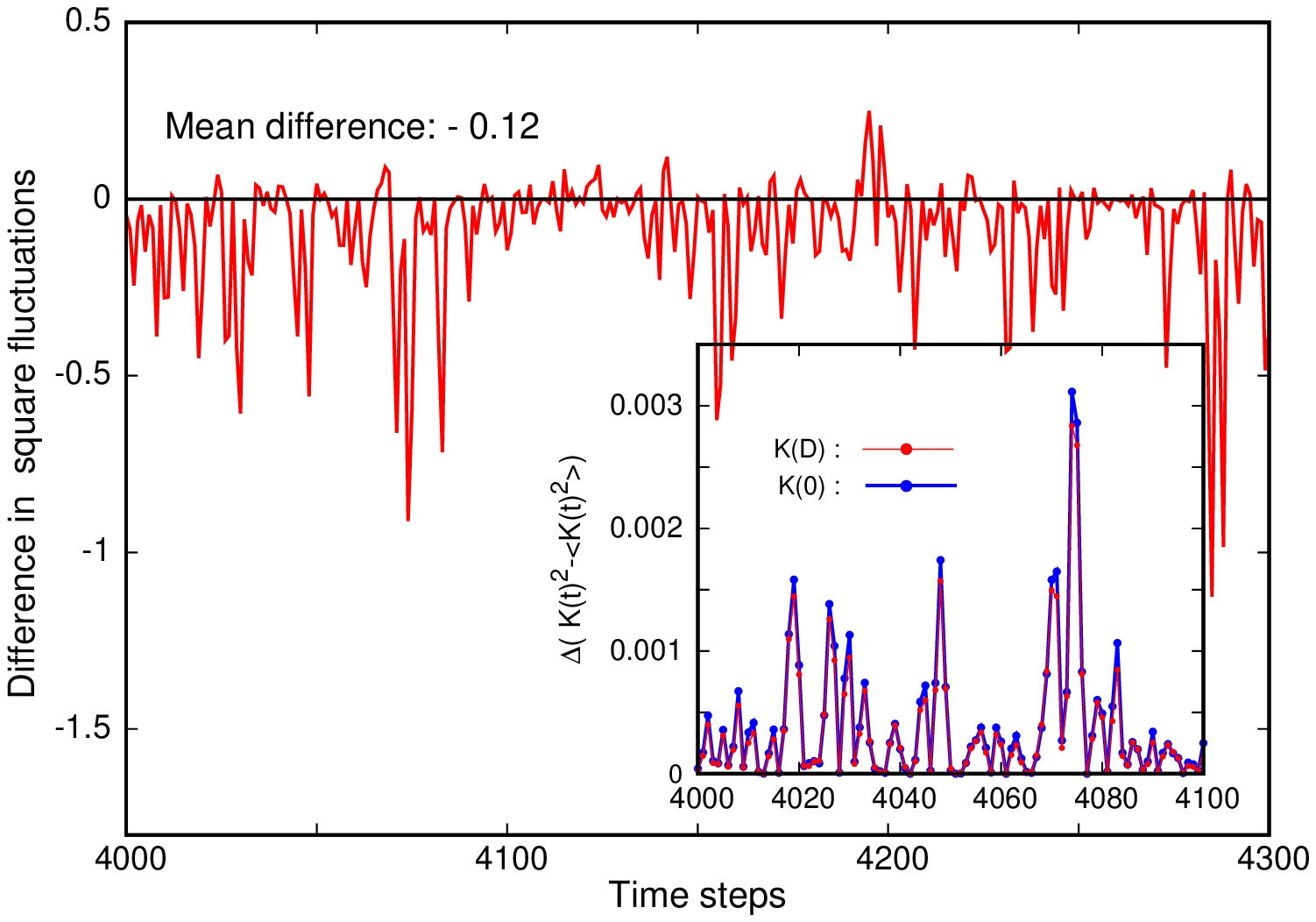}
		  \caption{ The relative difference in the time interval $t \in [4000 \delta t, 4300 \delta t]$   between the  fluctuations in the square of the excess  kinetic energies
		  $\Delta C_V^i$ for the LJ system at $(T,\rho)= (1.00, 1.40)$
		  and with $\delta t= 0.010$. The mean of the difference of the square excess kinetic energies  (Eq. 24) is -0.12.
		  The inset shows the two excess square kinetic energies per particle: with red is $K_D(t)^2-\bar{K}_D^2$ and with
		  blue is $K_0(t)^2-\bar{K}_0^2$.}
	  \end{center}		  
  \end{figure}
 \begin{figure}
	  \begin{center}	  
 	 \includegraphics[width=5.6cm,angle=-90]{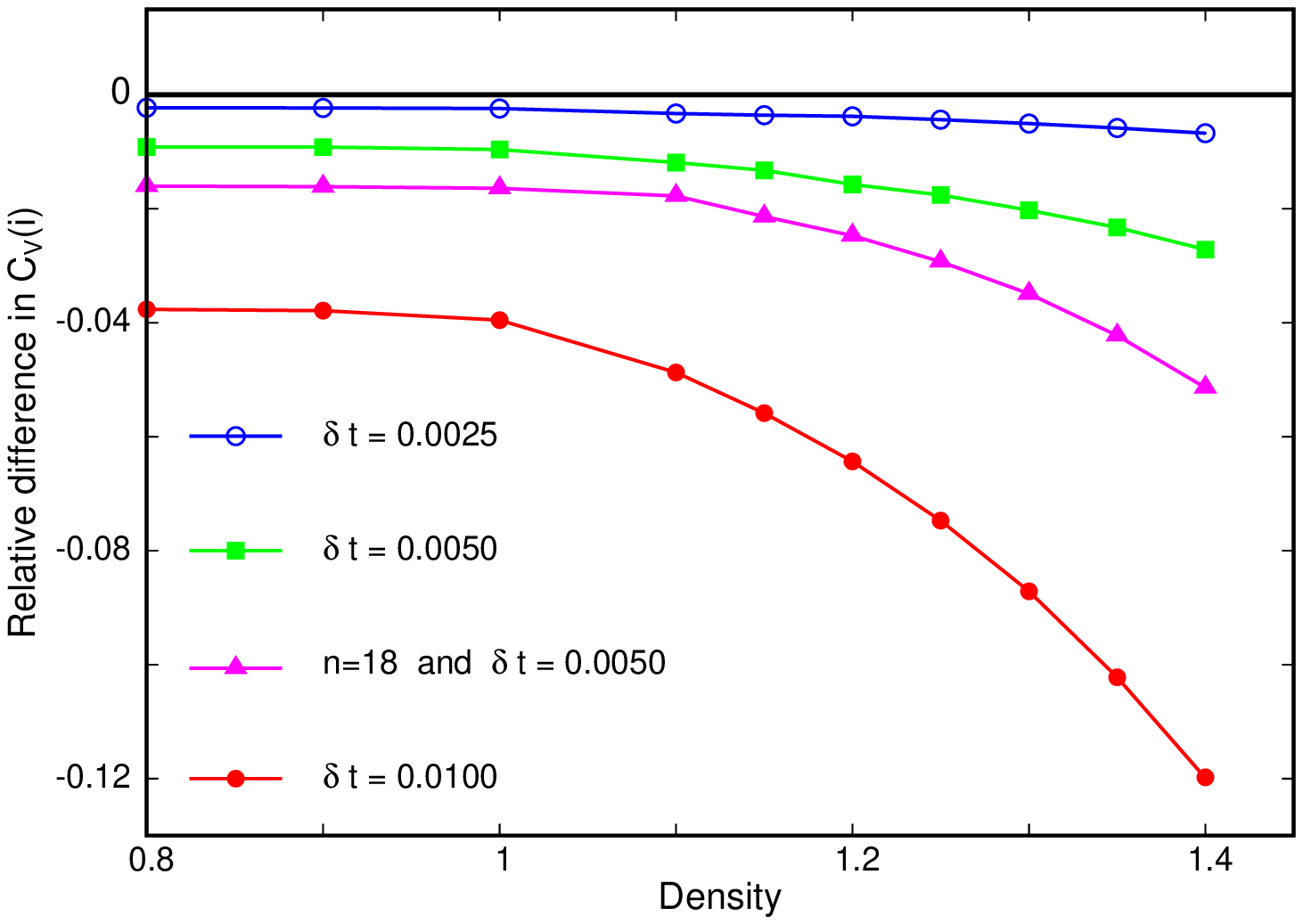}
		  \caption{ The relative difference between the  fluctuations in the square of the excess  kinetic energies
		  $\Delta C_V^i $ for the LJ system at $T= 1.00$
		  for different densities and time increments. The corresponding  relative differences in  temperatures are shown in Fig. 2.}
	  \end{center}		  
  \end{figure}

The  relative difference between fluctuations in the square of the excess  kinetic energies
   $\Delta C_V^i$ for a few hundred time steps  
   is shown in Fig. 4 for the LJ system, and with the corresponding  fluctuations  $K_D(t)^2-\bar{K}_D^2$  and  $K_0(t)^2-\bar{K}_0^2$ 
   in the inset in the  figure. The fluctuations are for the LJ system at $(T_D, \rho)=(1.00, 1.40)$ with the corresponding
   temperature fluctuations shown in Fig. 3. At first glance, the fluctuations are synchronous (Fig. 3 and the inset in Fig. 4), but
a closer examination reveals that there are significant differences and
of the order of twelve percent for the state point with relatively high density $\rho=1.40$ and
for the time increment $\delta t=0.010$. 

The relative difference in $C_V^i$ from $10^6$ time steps for the system at $T_D=1.00$ and at different densities and  time increments is shown in Fig. 5.
(The corresponding relative temperature differences are shown in Fig. 2.) The differences between $C_V^i$ obtained by fluctuations
in $K_D(t)$  and  $K_0(t)$, respectively, increases with 
increasing density, temperature, strength of the repulsive forces, or time increment.

\section{ Conclusion.}
Newton published his discrete algorithm in \textit{Principia Proposition I }, and the algorithm is now  
used everywhere in natural science disciplines ranging from simulations of planetary systems to simulations of
atoms, molecules, and systems with complex proteins. 
 L. Verlet rederived in 1967 the algorithm from a forward and backward Taylor expansion \cite{Toxvaerdappendix}
 and performed his MD simulations 
 with Eq. (5) for the velocity and Eq. (9) for the kinetic energy.  Discrete dynamics with Newton's algorithm 
has been further developed
 and is now a standard tool used in natural science, and MD was honored with a Nobel Prize in 2013.

Newton's discrete dynamics has the same properties as his analytic Classical Mechanics, it is time reversible, symplectic, and
 has the same invariances for a conservative system \cite{Toxvaerd2023}. Furthermore, there exists a shadow Hamiltonian
 where the positions of the discrete dynamics are located on the analytical trajectories for
 the shadow Hamiltonian \cite{Toxvaerd2023,Toxvaerd1994,Toxvaerd2012}. This means that there is no qualitative difference between
the analytical and the discrete dynamics, and MD with the Newton-Verlet algorithm is the exact generation of positions for the discrete
dynamics.

Verlet's approximative expression for  the velocity at time $t$, Eq. (5), which is used in  MD simulations
does not affect the discrete dynamics and the discrete positions, but it is an  unnecessary  approximation.
It is the first term in his Taylor expansion, 
and the corresponding expression for the kinetic energy, Eq. (9) ought to 
 be replaced with the exact expression, Eq. (14). Doing so one  achieves an agreement between the kinetic
 and  configurational temperature (Fig. 3). Potential energy fluctuations $\delta U$
 should be obtained by Eq. (11) or  (13), and velocities and their time correlations
 by the (constant) velocities in the respective time intervals.  Most MD simulations are canonical $NVT$ ensemble or $NPT$ ensemble simulations
 with the thermostat temperature $T(thermostat)=T_0$, and
not only are the $NVT_0$ and $NPT_0$ temperatures wrong so are also the values of $C_V^i$.
 The errors caused by using 
 Eq.(5) for the velocity and  Eq. (9) for the kinetic energy might only be a few percent for systems with relatively weak forces. But
the errors increase according to Eq. (22)  with the strength of the forces and the time increment $\delta t$ used in the simulations, and
can  lead to severe errors for systems with strong forces such as full-atomic models of e.g.
biomolecules with fast intramolecular vibrations.

The biggest challenge with an MD simulation of
a real system, however, is to formulate a correct force field,
and also for this reason   it is important to use the correct expression for
the kinetic energy, whereby one avoids introducing a systematic error in the calculated energies.
The public software used for MD simulations should be corrected.

\end{document}